\documentclass[12pt,a4paper,draft,reqno]{amsart}
\usepackage[english]{babel}
\usepackage{a4wide,amssymb,amsmath,xypic,ifthen}

\newcommand{\qee}{ \hfill\hspace{2pt}$\triangle$}
\newtheorem{thm}{Theorem}[section]

\theoremstyle{remark}
\newtheorem{rema}[thm]{Remark}
 
\newtheorem{exe}[thm]{Example}
 
\newcommand{\iso}{ \ \raise 4pt\hbox{$\sim$} \kern -10pt\hbox{$\to$}\  }

\newcommand{\M}{{\mathfrak M}}

\newcommand{\E}{{\mathcal E}}

\newcommand{\Z}{{\mathbb Z}}

\newcommand{\C}{{\mathbb C}}

\newcommand{\beq}{\begin{equation}}
\newcommand{\eeq}{\end{equation}}
\newcommand{\beqa}{\begin{eqnarray}}
\newcommand{\eeqa}{\end{eqnarray}}

\newcommand{\uno}{\mbox{1\kern-.59em {\rm l}}}

\def\val{{\vec{k}}}
\def\va{{\vec{a}}}

\newcommand{\be}{\begin{equation}}
\newcommand{\ee}{\end{equation}}
\newcommand{\bea}{\begin{eqnarray}}
\newcommand{\eea}{\end{eqnarray}}

\begin{document}
\begin{flushright} SISSA  28/2011/FM-EP
\end{flushright}
\title[Instantons on ALE spaces and Super Liouville]{Instantons on ALE spaces and Super Liouville Conformal Field Theories}
\bigskip
\date{\today}
\thanks{This research was partly supported by the INFN Research Project PI14 ``Nonperturbative dynamics of gauge theory", by   PRIN    ``Geometria delle variet\`a
algebriche" and by the INFN Research Project TV12.
 \\[5pt] \indent e-mail: {\tt bonelli,maruyosh,tanzini @sissa.it} }
 \maketitle \thispagestyle{empty}
\begin{center}{\sc Giulio Bonelli},  {\sc Kazunobu Maruyoshi} and {\sc Alessandro Tanzini}
\\[10pt]  {\small
 Scuola Internazionale Superiore di Studi Avanzati, \\ Via Bonomea 256, I-34136
Trieste, Italia \\ and \\ Istituto Nazionale di Fisica Nucleare, Sezione di Trieste.
}
\end{center}

\begin{abstract} 
We provide evidence that the conformal blocks of ${\mathcal N}=1$ super Liouville conformal field theory
are described in terms of the $SU(2)$ Nekrasov partition function on the ALE space ${\mathcal O}_{\mathbb{P}^1}(-2)$.
  \end{abstract}

\bigskip
\section{Introduction}

In \cite{AGT} it was proposed that the conformal block of Liouville conformal field theory (CFT) are directly related
to the instanton Nekrasov partition function of quiver ${\mathcal N}=2$ superconformal gauge theories on $\mathbb{C}^2$.
This correspondence was extended in \cite{gaiotto} to asymptotically free theories whose Nekrasov partition functions
are shown to be equal to the norm of suitable Whittaker vectors. 

More recently, in \cite{BF} an intriguing observation has been put forward, 
stating that conformal blocks in ${\mathcal N}=1$ super Liouville theory 
can be reproduced by a $\mathbb{Z}_2$ projection of the instanton counting on $\mathbb{C}^2$.
It was shortly later proposed in \cite{NT} that 
counting 
$N$ M5 branes on $\mathbb{C}^2/\mathbb{Z}_m$ in presence of an $\Omega$-background 
realize para-Liouville/Toda CFTs. This reduces in the $m=2$ case to the ${\mathcal N}=1$ super Liouville CFT.
Evidence in this direction is obtained by computing the relevant central charges generalizing the approach proposed in 
\cite{BT,ABT}.

In this letter we analyze the four dimensional gauge theory side of the correspondence proposing that
the relevant gauge theories are to be formulated on ALE spaces, 
as it is suggested by the M-theory construction. 
We provide a check of this proposal by comparing the Nekrasov partition function on $\mathbb{C}^2/\mathbb{Z}_2$,
whose minimal resolution is the Eguchi-Hanson space $T^*\mathbb{P}^1={\mathcal O}_{\mathbb{P}^1}(-2)$,
with the super Liouville conformal blocks.

\section{Instantons on ALE spaces}

The moduli spaces of anti-self-dual connections on ALE spaces $\mathbb{C}^2/\Gamma$ with Kleinian subgroup $\Gamma\in SU(2)$
where described in \cite{KN} in terms of representations of quivers associated to the ADE (extended) Dynkin diagrams,
which provide a parametrization of these moduli spaces in terms of ADHM data.
Using these results and the localization techniques on the instanton moduli spaces \cite{Nekra,Flupo,BFMT} a characterization of 
the fixed points under the relevant torus action where provided and used
to evaluate partition functions of supersymmetric gauge theories on $\C^2/\Z_n$ in \cite{Fucito:2004ry}.
Here we will follow a different approach based on the algebraic geometry description of the same data. This is 
obtained \cite{nakamoscow} by considering the moduli space of framed torsion free sheaves on the global quotient $\mathbb{P}^2/\Gamma$
with minimal resolution of the singularity at the origin. For $\Gamma=\mathbb{Z}_2$
the resulting variety 
corresponds to a ``stacky" compactification $\mathcal X_2$ of ${\mathcal O}_{\mathbb{P}^1}(-2)$ obtained by adding a divisor 
$\tilde C_\infty\simeq \mathbb{P}^1/\mathbb{Z}_2$ \cite{bpt}.  
The fixed points under the torus action of this moduli
space where classified in \cite{bpt}\footnote{In \cite{bpt} a general analysis of the total spaces of 
${\mathcal O}_{\mathbb{P}^1}(-p)$ were performed. Let us notice that for $p\ne2$ the 
quotient considered in \cite{bpt} is not of ADE type, and it would be interesting to investigate its 2d CFT counterpart.}.
The advantage of this approach is twofold: first of all it allows to easily recognize a kind of 
blow-up formula relating the Nekrasov partition function on ${\mathcal O}_{\mathbb{P}^1}(-2)$ to the standard
one on $\mathbb{C}^2$ -- see Eq.(\ref{zale}) -- similar to the ones found in \cite{NY-I,NY-L} for the blow-up of $\mathbb{P}^2$ . 
In this sense, it also considerably simplifies the combinatorics needed  
to perform the actual computations. 
 
Let us recall the classification of the fixed points of \cite{bpt}, to which we refer for details\footnote{See also \cite{sasaki}.}.
The moduli space $\widetilde\M^2(r,k,n)$ of framed torsion free coherent sheaves on $\mathcal X_2$ is characterized 
by the rank $r$, the first Chern
class $c_1(\mathcal{E})= k C$, where $C$ is the exceptional divisor resolving the singularity at the origin, and the discriminant
$$ \Delta(\E)=c_2(\E)-\frac{r-1}{2r}c_1^2(\E)= n \ .$$
Let us remark that $k$ is in general half integer due to the ``stacky'' compactification of the ALE space.
Indeed, half-integer classes take into account anti-self-dual connections which asymptote flat connections with
non trivial holonomy at infinity.

The torus action we consider is the standard action of the Cartan torus of the gauge group 
parametrized in terms of the scalar vevs $\vec{a}=\{a_\alpha\}$, $\alpha=1,\ldots,r$ times the 
space-time rotations acting on ${\mathcal O}_{C}(-2)$ as $T: [z:w]\to [t_1z:t_2w]$ on the exceptional divisor $C$ and as 
$(z_1,z_2)\to (t_1^2 z_1, t_2^2 z_2)$
on the fibers over it. 
The fixed points of $\widetilde\M^2(r,k,n)$ under this action are labeled in terms of $\vec{k}=(k_1,\ldots, k_r)$ and 
of two sets of Young tableaux $(\vec{Y}^1,\vec{Y}^2)$,
where $\vec{k}$ parametrize the first Chern class as $\sum_{\alpha=1}^r k_\alpha=k$
and $\vec{Y}^1=\{ Y^1_\alpha\}$, $\alpha=1,\ldots,r$ parametrize the ideal sheaves
supported at $w=0$, while $\vec{Y}^2=\{ Y^2_\alpha\}$, $\alpha=1,\ldots,r$ the ones supported at $z=0$.
These are the two fixed points of the exceptional divisor $C$ under the torus action $T$.
The above data are constrained by the relations
\bea
n &=&  \sum_\alpha  \left( \vert
Y_\alpha ^1 \vert +\vert Y_\alpha ^2 \vert \right) +
\frac{1}{r} \sum_{\alpha <\beta} (k_\alpha  -k_\beta)^2\, , \nonumber\\
k&=&\sum_{\alpha=1}^r k_\alpha
\label{countboxes}
\eea
Looking for all collections of Young tableaux and strings of integers $\vec{k}$ satisfying these conditions 
one enumerates all the fixed points. 

The tangent space at the fixed points has the following weight decomposition with respect to the torus action
\bea
\label{tangent}
&&   T \, \widetilde\M^2(r,k,n)
   = \\ \nonumber
 &&\sum_{\alpha ,\beta=1}^r \left(L_{\alpha ,\beta}(t_1,t_2)
     + t_1^{2(k_\beta - k_\alpha )} N_{\alpha ,\beta}^{\vec{Y}_1}(t_1^2,t_2/t_1)
     + t_2^{2(k_\beta - k_\alpha )}
     N_{\alpha ,\beta}^{\vec{Y}_2}(t_1/t_2,t_2^2)\right),\label{tangent_space}
\eea 
where $L_{\alpha ,\beta}(t_1,t_2)$ is given by 
\begin{equation}
L_{\alpha, \beta} (t_1,t_2) 
=  e_\beta\, e_\alpha ^{-1}
\sum_{{i,j\ge 0, i+j-2n_{\alpha\beta} \equiv 0 \ {\rm mod} 2, \ i+j \le 2(n_{\alpha\beta}-1)}}
      t_1^{-i} t_2^{-j} 
\label{Lmag0}
\end{equation}
for $n_{\alpha\beta}\equiv k_\alpha - k_\beta>0$ and by 
\begin{equation}
L_{\alpha, \beta} (t_1,t_2) 
 =   e_\beta\,e^{-1}_\alpha
\sum_{{i,j\ge 0,\ i+j+2+2n_{\alpha\beta}\equiv 0 \ {\rm mod} 2, \ i+j \le - 2n _{\alpha\beta}- 2}}
      t_1^{i+1} t_2^{j+1} 
\label{Lmin0}
\end{equation}
for  $n_{\alpha\beta}<0$.
Finally
 \be
N_{\alpha ,\beta}^{\vec{Y}}(t_1,t_2)=e_\beta e_\alpha ^{-1}\times
\left\{\sum_{s \in Y_\alpha }
\left(t_1^{-l_{Y_\beta}(s)}t_2^{1+a_{Y_\alpha }(s)}\right)+\sum_{s
\in Y_\beta}
\left(t_1^{1+l_{Y_\alpha }(s)}t_2^{-a_{Y_\beta}(s)}\right)\right\}\,,
\label{R4_tangent_space} \ee
namely it coincides with the weight decomposition of the tangent of the moduli space on $\C^2$. 
Here $\vec Y$ denotes an $r$-ple  of Young tableaux, while for a given box $s$ in the tableau
$Y_\alpha$, the symbols $a_{Y_\alpha}$ and $l_{Y_\alpha}$ denote the ``arm" and ``leg" of $s$ respectively, that is, the number of boxes above and on the right to $s$.

By using the above results one can readily compute the 
the instanton part of the Nekrasov partition function on $\mathcal{O}_{\mathbb{P}^1}(-2)$. 
In order to compare with the super Liouville conformal blocks,
we are interested in the case of $SU(2)$ gauge group. In this case $c_1=0$ mod $2$, while the discriminant
$n$ coincides with the second Chern character $ch_2(\mathcal{E})$.
Moreover $\vec{a}=(a,-a)$.
We resum the contributions at fixed $ch_2=n$ against the instanton 
topological action $q^{ch_2}=q^n$. Moreover, since (\ref{tangent}) and $n$ both depend only on the difference $k_1-k_2=n_{12}$,  
we pick a single representative 
$[\val]$
for each class of 
$\val$  at fixed topological classes (\ref{countboxes}).
The instanton partition function then reads
\bea
\label{zale}
&&{\mathcal Z}_{inst}^{ALE}(q,\epsilon_1,\epsilon_2,\va)=\\
&&
\sum_{[\val] | c_1=0 \, {\rm mod}\, 2} 
\frac{q^{
\frac{1}{2}(n_{12})^2}}
{\prod_{\alpha,\beta} l^{\val}_{\alpha\beta}\left(\epsilon_1,\epsilon_2, \va \right)}
Z_{inst} \left(2\epsilon_1,\epsilon_2-\epsilon_1,\va+2\epsilon_1\val,q\right)
Z_{inst}\left(\epsilon_1-\epsilon_2,2\epsilon_2,\va+2\epsilon_2\val,q\right)
\nonumber
\eea
where
$Z_{inst}\left(\epsilon_1,\epsilon_2,\va,q\right)$
is the Nekrasov partition function
on $\C^2$ and
$ l^{\val}_{\alpha\beta}\left(\epsilon_1,\epsilon_2, \va \right)$ is the product of the eigenvalues
of $L_{\alpha,\beta}(t_1,t_2)$ in the weight decomposition of the tangent space (\ref{tangent}).

\section{Comparison with super Liouville CFT}
  In this section, we check that 
  the instanton partition function of $\mathcal{N}=2$ pure $SU(2)$ gauge theory 
  on the minimal resolution of $\mathbb{C}^2/\mathbb{Z}_2$
  agrees with the norm of the Whittaker vector of Super Virasoro algebra as constructed in \cite{BF}.
  
  Firstly we compute the factor in the denominator in (\ref{zale}).
  In the $SU(2)$ case, this factor is just
  $l_{12}^{\vec{k}}(\epsilon_1, \epsilon_2, \vec{a}) l_{21}^{\vec{k}}(\epsilon_1, \epsilon_2, \vec{a})$
  and depends on the difference $n_{12}$.
  For simplicity, let us list
  the $l_{\alpha \beta}^{\vec{k}}$ for $n_{12} = 0, 1, 2$ which are the values needed in our computation below:
    \bea
    n_{12} =0;
    & &    l_{\alpha \beta}^{\vec{k}} = 1,
           \nonumber \\
    n_{12} =1;
    & &    l_{12}^{\vec{k}} = -2a, ~~~
           l_{21}^{\vec{k}} = 2a + \epsilon_1 + \epsilon_2
           \nonumber \\
    n_{12} =2;
    & &    l_{12}^{\vec{k}} = -2a (-2a - \epsilon_1 - \epsilon_2)(-2 a - 2 \epsilon_1)(-2a - 2 \epsilon_2)
           \nonumber \\
    & & 
           l_{21}^{\vec{k}} = 
           (2a + \epsilon_1 + \epsilon_2)(2a + 2\epsilon_1 + 2 \epsilon_2)
           (2a + 3 \epsilon_1 + \epsilon_2)(2a + \epsilon_1 + 3 \epsilon_2).
           \label{k1221}
    \eea
  For $n_{12}<0$, $l_{\alpha \beta}$ is obtained 
  by exchanging the indices $1$ and $2$ and $a \rightarrow -a$ in the above formulas.
  
  Let us calculate the instanton expansion of the partition function:
    \bea
     {\mathcal Z}_{inst.}^{ALE}(q,\epsilon_1,\epsilon_2,\va)
     =     \sum_{n=0}^\infty q^{n} {\mathcal Z}_n,
    \eea
  where $n= |\vec{Y^1}| + |\vec{Y^2}| + \frac{1}{2} n_{12}^2$.
  At the level of $n = \frac{1}{2}$, the only possible choice is $n_{12} = \pm 1$ and no box.
  Therefore, by using (\ref{k1221}) we obtain
    \bea
    {\mathcal Z}_{1/2}
     =   - \frac{2}{(2 a+ \epsilon_1 + \epsilon_2)(2a - \epsilon_1 - \epsilon_2)}.
    \eea
  When $n=1$, the only possibility is $n_{12}=0$ and $(|\vec{Y^1}|, |\vec{Y^2}|) = (1,0)$ or $(0,1)$.
  This can easily evaluated from the sum of two one-instanton contributions on $\mathbb{C}^2$ 
  whose arguments are shifted as in (\ref{zale}) respectively: 
    \bea
    {\mathcal Z}_{1}
     =   - \frac{1}{\epsilon_1 \epsilon_2(2 a+ \epsilon_1 + \epsilon_2)(2a - \epsilon_1 - \epsilon_2)}.
    \eea
  At $n=\frac{3}{2}$, $n_{12}= \pm 1$ and $(|\vec{Y^1}|, |\vec{Y^2}|) = (1,0)$ or $(0,1)$.
  The partition function is computed as
    \bea
    {\mathcal Z}_{3/2}
    &=&    \frac{1}{\epsilon_1 \epsilon_2
           (2a + \epsilon_1 + \epsilon_2)
           (2a - \epsilon_1 - \epsilon_2)}
           \nonumber \\
    & &    ~~~~~\times
           \frac{2 (4 a^2 - 22 \epsilon_1 \epsilon_2 - 9 \epsilon_1^2 - 9 \epsilon_2^2)}{
           (2a - \epsilon_1 - 3\epsilon_2)(2a - 3\epsilon_1 - \epsilon_2)
           (2a + \epsilon_1 + 3\epsilon_2)(2a + 3\epsilon_1 + \epsilon_2)}.
    \eea
  At $n=2$, there are two cases; $n_{12} = \pm 2$ with no box and $n_{12} = 0$ 
  with $(|\vec{Y^1}|, |\vec{Y^2}|) = (2,0), (1,1), (0,2)$:
    \bea
    {\mathcal Z}_{2}
    &=&    \frac{16a^4 - 52 a^2 \epsilon_2^2 - 92a^2 \epsilon_1 \epsilon_2 - 52a^2 \epsilon_1^2
         + 177\epsilon_1^3 \epsilon_2 + 177 \epsilon_1 \epsilon_2^3 + 36\epsilon_1^4 + 294 \epsilon_2^2 \epsilon_1^2
          +36 \epsilon_2^4}{8 \epsilon_1^2 \epsilon_2^2 (a + \epsilon_1 + \epsilon_2)(a - \epsilon_1 - \epsilon_2)
          (2a - \epsilon_1 - \epsilon_2)(2a + \epsilon_1 + \epsilon_2)}
           \nonumber \\
    & &    ~~~~~\times
          \frac{1}{(2a + 3 \epsilon_1 + \epsilon_2)(2a + \epsilon_1 + 3 \epsilon_2)
          (2a - \epsilon_1 - 3 \epsilon_2)(2a - 3 \epsilon_1 - \epsilon_2)}.
    \eea
  It is easy to extend this calculation to higher orders. 
  
  We compare these instanton coefficients with the norm of the Whittaker vector as given in \cite{BF}.
  The conformal blocks in the Whittaker limit are 
    \be
    F_0 \left(q,c,\Delta\right)
     =     \sum_{n\in{\mathbb N}} q^n B_n
    \ee
  in the Neveu-Schwarz sector and
    \be
    F_1 \left(q,c,\Delta\right)
     =     \sum_{n\in{\mathbb N}+\frac{1}{2}} q^n B_n
    \ee
  in the Ramond sector.
  For comparison, we list the results
    \bea
    B_{1/2}
    &=&    \frac{1}{2 \Delta}, ~~~
    B_1
     =     \frac{1}{8 \Delta}, ~~~
    B_{3/2}
     =     \frac{c + 2 \Delta}{8 \Delta (c - 6 \Delta + 2c \Delta + 4 \Delta^2)},
           \nonumber \\
    B_2
    &=&    \frac{3 c + 3 c^2 - 34 \Delta + 22 c \Delta + 32 \Delta^2}{64 \Delta (-3 + 3c + 16 \Delta)
           (c - 6 \Delta + 2c \Delta + 4 \Delta^2)}, ~~\ldots
    \eea
  with the conformal dimension $\Delta = \frac{(b + 1/b)^2}{8} - \frac{\lambda^2}{2}$
  and the central charge $c = 1 + 2 (b + 1/b)^2$.
  These agree with the instanton coefficients by the following relations
    \bea
    {\mathcal Z}_{n}
    &=&    B_{n}, ~~~{\rm if}~n \in \mathbb{N}
           \nonumber \\
    {\mathcal Z}_{n}
    &=&    \frac{B_{n}}{2}, ~~~{\rm if}~n \in \mathbb{N} + \frac{1}{2},
    \eea
  under the identification of the parameters \cite{BF}:
    \bea\label{id}
    a
     =     \lambda, ~~~~
    \epsilon_1
     =     b, ~~~
    \epsilon_2
     =     \frac{1}{b}.
    \eea
  Note that on the gauge theory side the parameters have been understood as dimensionless.
  We have checked these relations up to $n = \frac{5}{2}$, but we omit the explicit result for brevity.

The outcome of the above, is summarized by the equality
\be
{\mathcal Z}_{inst.}^{ALE}(q,\epsilon_1,\epsilon_2,\va)=F_0\left(q,c,\Delta\right)+\frac{1}{2}F_1\left(q,c,\Delta\right)
\label{final}\ee
under the identification of the parameters (\ref{id}).
The complete proof of (\ref{final}) should pass by a full characterization of the analytic structure of the two sides
which could be considerably simplified by using eq.(\ref{zale}).

Eq.(\ref{final}) in particular identifies the sector with trivial boundary conditions 
in the gauge theory with the even sector in super Liouville theory 
and the sector with non-trivial boundary conditions in the gauge theory, namely flat connections with  
non trivial holonomy, with the odd sector in the super Liouville theory.
Notice the absence of any $U(1)$ factor in the above identification due to the fact that we are dealing with 
the pure $SU(2)$ gauge theory.

\section{Conclusions and discussions}

There are various issues raised by our analysis one can start discussing.
The most relevant coming to our minds are the followings.

The blow-up formula (\ref{zale}) indicates a precise relation between Virasoro and super Virasoro conformal blocks, once
eq.(\ref{final}) and the equivalence of the Nekrasov partition function on ${\mathbb C}^2$ with standard Virasoro conformal blocks 
\cite{AGT} are used. 
The meaning of these relations in CFT should be understood.

Our analysis should be completed by comparing the perturbative part of the Nekrasov partition function on ALE spaces 
with the three point functions of the super Liouville theory \cite{RS,P}.
This could open the possibility of studying correlators in the super Liouville theory in terms of $\mathcal{N}=2$
gauge theory on $S^4/\Gamma$, by generalizing the approach of \cite{Pestun} and its CFT interpretation.

Moreover, one could consider super Liouville theory on more general Riemann surfaces
which should be related with quiver gauge theories on ALE space.
The case with matter in the fundamental representation will be presented elsewhere \cite{wip}.

Furthermore, our result opens the way to generalize the AGT correspondence to all the 
ADE quotients of ${\mathbb C}^2$ by comparing the instanton counting with paraLiouville/Toda
theories as suggested in the M-theory construction in \cite{NT}.

It would indeed be most interesting to further analyze the geometry of the M-theory compactification
for this class of theories in order to realize if the low energy dynamics of the gauge theory on ALE space can be encoded in the 
spectral geometry of some integrable system, presumably related to the Hitchin system. 
Indeed a parallel between instanton moduli space on ALE and Hitchin systems was pointed out in \cite{naka-ale}.
According to the standard AGT-like approach,
the CFT counterpart should be encoded in the classical limit of stress-energy tensor or degenerate fields insertions. 
Actually,
it would be interesting to try to extend the relation we found in presence of surface oparators
which might correspond to the insertion of degenerate fields in super Virasoro conformal block as in \cite{AGGTV}.
Furthermore, this also enables us to extract the information of integrable systems related with the gauge theories on ALE spaces
by using the method in \cite{Teschner, AT, MT, MMM, BMT}.

The relation considered in this paper might be explained in terms of topological string theory.
One possibility to study this could pass by introducing matrix models associated with the conformal blocks as in \cite{DV}.
See also \cite{Kimura} for the matrix representation of the instanton partition function on ALE space.

The emergence of infinite dimensional Lie algebras was observed in the study of the cohomology of the instanton moduli spaces 
in \cite{naka-ale}. 
This allowed to uncover a relation of the $\mathcal{N}=4$ gauge theory partition function with 
the characters of affine Lie algebras, further studied in \cite{VW,Dijkgraaf1, Dijkgraaf2}.
This could be related to the correspondence that we obtain/suggest in this paper.
Indeed, let us notice that, by including a further parameter $y$ sourcing the $c_1$ coupling in the gauge theory 
generating function, the right hand side of equation (\ref{zale}) gets an extra factor of
$\sum_{c_1\,\,{\rm even}} y^{c_1} q^{{c_1}^2/8}=\vartheta_3\left(y,q\right)$.

\vspace{1cm}
{\bf Acknowledgments}: It is a pleasure to thank T.~Eguchi, T.~Nishioka, R.~Rashkov and Y.~Tachikawa
for useful discussions.

\par

\end{document}